\documentclass{cta-author}

{}
{}
{}

\usepackage{subcaption} 
\usepackage[exponent-product=\ensuremath{\cdot}]{siunitx}
\usepackage{mathtools}
\usepackage{tikz, pgfplots}
\usepackage{tudacolors}
\usepackage{pgfplotstable}
\usepackage{pdfcomment}
\usepackage{tabularx}

\newcommand{\R}{\mathbb{R}}

\newcommand{\curl}{\operatorname{curl}}
\newcommand{\param}{\ensuremath{\mathbf{p}}}
\newcommand{\stiff}{\ensuremath{\mathbf{K}}}
\newcommand{\mass}{\ensuremath{\mathbf{M}}}

\newcommand{\nvar}{n}
\newcommand{\paramn}{\ensuremath{{p_{\nvar}}}}
\newcommand{\paramnabbrev}{\ensuremath{p}}
\newcommand{\degr}{\ensuremath{d}}

\renewcommand{\P}{\mathbf{P}}
\newcommand{\ddt}{\frac{\mathrm{d}}{\mathrm{d}\paramnabbrev}}
\newcommand{\ddp}{\frac{\mathrm{d}}{\mathrm{d}\param}}
\def\Rtune{\ensuremath{R_{\mathrm{arc}}}}

\begin{document}

\supertitle{Submission Template for IET Research Journal Papers}

\title{Gradient-Based Eigenvalue Optimization for Electromagnetic Cavities with Built-in Mode Matching}

\author{\au{Anna Ziegler$^{1\corr}$}, \au{Robert Hahn$^{1}$}, \au{Victoria Isensee$^{1}$, \au{Anh Duc Nguyen$^{1}$}, \au{Sebastian Schöps$^{1}$}}}
\address{\add{1}{Computational Electromagnetics Group, Technische Universität Darmstadt, Schloßgartenstr. 8,
64289 Darmstadt, Germany}
\email{anna.ziegler@tu-darmstadt.de}}

\begin{abstract}
Shape optimization with respect to eigenvalues of a cavity plays an important role in the design of new resonators or in the optimization of existing ones.
In our paper, we propose a gradient-based optimization scheme, which we enhance with closed-form shape derivatives of the system matrices.
Based on these, we can compute accurate derivatives of eigenvalues, eigenmodes and the cost function with respect to the geometry, which significantly reduces the computational effort of the optimizer.
We demonstrate our work by applying it to the 9-cell TESLA cavity, for which we tune the design parameters of the computational model to match the design criteria for devices in realistic use cases.
Since eigenvalues may cross during the shape optimization of a cavity, we propose a new algorithm based on an eigenvalue matching procedure, to ensure the optimization of the desired mode in order to also enable successful matching along large shape variations.
\end{abstract}

\maketitle

\section{Introduction} \label{introduction}
Shape optimization of eigenvalue problems is a challenging task and has been addressed with different approaches.
One option is the use of evolutionary algorithms which can provide simple implementations and sufficiently good solutions~\cite{Brackebusch_2013aa, Kranjcevic_2019aa, Kranjcevic_2019ab, Udongwo_2023aa}. 
However, these methods are often inefficient as they select the best solution obtained by evaluating many points in the parameter space \cite{Kennedy_1995aa, MezuraMontes_2011aa, Pedersen_2010aa}.
On the other hand, in order to beneficially employ gradient-based optimization approaches, analytical derivatives or sufficiently good approximations of these are lacking in many settings. 
The work of~\cite{Valles_2009aa} circumvents this issue by employing a gradient-free optimizer implemented in MATLAB\textsuperscript{\textregistered}’s function \texttt{fminsearch} which requires them to reformulate their problem as an unconstrained program by penalizing deviations from the physical requirements.
Other derivative-free approaches are presented in~\cite{Hassan_2015aa} and \cite[Ch. 6.4]{Corno_2017ad}, for which the authors use a trust region optimization method relying on successively updating surrogate models. 
The adjoint method used in~\cite{Herter_2023aa, Toader_2017aa, Akelik_2005aa} is another powerful tool, especially when there a many design variables. 
Alternatively, in \cite{Putek_2022aa}, the authors applied shape derivatives of functionals in the continuous framework using the velocity and adjoint variables for an enhancement of the steepest descent algorithm in a stochastic setting.

In our work, we consider a shape optimization for which we apply closed-form shape derivatives of the eigenvalue and eigenmodes.
This formulation can be used in a variety of applications, such as the design of new cavity geometries or in the optimization of existing ones.
As the cavity's eigenvalues, i.e., frequencies, depend sensitively on the geometry of the structure, geometry parameters need to be determined carefully in order to achieve the required resonant frequency and flatness of the electric field, i.e., the even distribution of the electric field throughout the cavity.
Here, we want to exemplify our algorithm by applying it to tune a given model of a cavity in order to match it to tuned devices from practice with respect to its field patterns.

The paper is structured as follows. 
In Section \ref{problem}, we state the Maxwell eigenvalue problem as well as its discretized counterpart and give an initial formulation of the optimization problem.
Based on the isogeometric discretization, we demonstrate the derivation of closed-form shape derivatives in Section~\ref{sec:iga_matrix_diff}.
In Section~\ref{crossing}, we propose an extension for treating the crossing of the eigenvalues along the optimization. 
Numerical examples are discussed in Section~\ref{numerics}, where we formulate more specific objective functions and evaluate the performance of our approach.
We conclude our work in Section~\ref{conclusion}.

\section{Problem formulation} \label{problem}
Starting from Maxwell's equations and assuming time-harmonic quantities, a non-conductive domain $\Omega_\param$ parametrized with a parameter vector~$\param$ and perfect electric conductor (PEC) boundary conditions on~$\partial \Omega_\param$, we recover the well-known wave equation. 
Formulated as an eigenvalue problem, it reads: Find all eigenpairs $\lambda = k^2 \in \R^+$ and $\mathbf{E} \in H_0(\curl)$ s.t.
\begin{equation}
\begin{alignedat}{2}
	\curl \left(\curl \mathbf{E} \right) &= \lambda \mathbf{E}\quad &&\mathrm{in }\, \Omega_\param \, , \\
	\mathbf{E} \times \mathbf{n} &= \mathbf{0}\quad &&\text{ on }\, \partial \Omega_\param \, ,
\end{alignedat}
\label{eq:eigenvalue-problem-maxwell}
\end{equation}
where $k = \omega \sqrt{\mu \varepsilon}$ is the wave number, $\mu$ and $\varepsilon$ are the permeability and permittivity within the domain, which we assumed to be filled with vacuum. 
The normal vector $\mathbf{n}$ is oriented outwards, and $H_0(\curl)$ 
contains square-integrable functions, for which the curl exists in a weak sense, and whose trace vanishes on the boundary. 
The corresponding weak formulation reads: Find all eigenpairs $\lambda \in \R^+$ and $\mathbf{E} \in H_0(\curl)$ s.t.		 
\begin{equation}
	\begin{alignedat}{2}
		\left< \curl \mathbf{E}, \curl \mathbf{v} \right> &= \lambda \left< \mathbf{E}, \mathbf{v} \right> \quad \forall \mathbf{v} \in H_0(\curl) \, , \\
	\end{alignedat}
	\label{eq:eigenvalue-problem-weak}
\end{equation}
and is obtained via the ($L^2$) inner product $\left<\cdot,\cdot\right>$ with tests functions $\mathbf{v}$, which are chosen from the same function space as the unknown~$\mathbf{E}$ as per the Ritz-Galerkin method~\cite{Monk_2003aa}.
Approximating $\mathbf{E}$ as $\mathbf{E} = \sum_{j=1}^{n_\mathrm{dof}} e_j \mathbf{v}_j$ with a finite number $n_\mathrm{dof}$ of coefficients and basis functions and arranging those coefficients $e_j$ into the vector $\mathbf{e}$ of degrees of freedom (DoF) leads us to the discrete generalized eigenvalue problem: Find all eigenpairs $\lambda \in \R^+$ and $\mathbf{e} \in \R^{n_\mathrm{dof}}$ s.t.
\begin{equation}
	\begin{alignedat}{2}
		\stiff\mathbf{e} &= \lambda \mass \mathbf{e}
	\end{alignedat}
	\label{eq:eigenvalue-problem-discrete}
\end{equation}
where the matrices are given by
\begin{equation}
	\begin{alignedat}{1}
		\stiff_{ij}(\param) &= \int_{\Omega_\param} \curl \mathbf{v}_i \cdot \curl \mathbf{v}_j \,\mathrm{d}\mathbf{x}\;, \\
		\mass_{ij}(\param) &= \int_{\Omega_\param} \mathbf{v}_i \cdot \mathbf{v}_j \,\mathrm{d}\mathbf{x}\;.
	\end{alignedat}
	\label{eq:system-matrices}
\end{equation}
Solving \eqref{eq:eigenvalue-problem-discrete} yields the eigenmodes $\mathbf{e}_k(\param)$ and eigenvalue $\lambda_k(\param)$ for each value of $\param$, and the eigenvalue is related to the frequency via
\begin{equation}
	f_k(\param) = \frac{\sqrt{\lambda_k(\param)}}{2 \pi \sqrt{\mu \varepsilon}} \text{.}
	\label{eq:relation-lambda-f}
\end{equation}
Since the ordering of eigenvalues and eigenfrequencies will be relevant in the following, we always assume that they are sorted according to the frequency, i.e., 
\begin{equation}
    f_1\leq f_2 \leq \ldots \leq f_k \leq \ldots \leq f_{n_\mathrm{dof}}
	\label{eq:sort}
\end{equation}
for each value of $\param$.  

\subsection{Formulation of the Optimization Problem}
In our eigenvalue optimization setting, we want to determine the appropriate $\param$ to achieve that the $k$-th eigenvalue $\lambda_k$ equals some given $\lambda_\mathrm{ref}$, where we disregard eigenvalue crossings for now.
With the basic squared-error cost function
\begin{equation}
	g(\lambda_k(\param)) = \frac{1}{2}\left(\lambda_\mathrm{ref}- \lambda_k(\param) \right)^2 \mathrm{, }
	\label{eq:cost-fct}
\end{equation}
we can formulate the optimization problem
\begin{subequations}\label{eq:opti_problem}
\begin{alignat}{2}
	& \! \!  \min_{\mathbf{p}} \quad   & g(\lambda_k(\param))\\
    & \text{s.t.} \quad         &  \stiff(\param)\mathbf{e}_k(\param)   & = \lambda_k(\param) \mass(\param) \mathbf{e}_k(\param) \, , \label{eq:constraintGEVP}\\
    &                           & \mathbf{e}_{\star}^\top \mass(\param) \mathbf{e}_k(\param) &  = 1 \, , \label{eq:constraintNormalEigvec}\\
    &                           & 0\leq \paramn & \leq 1  \qquad \qquad \text{for all }\nvar \, ,
\end{alignat}
\end{subequations}
where we have assumed w.l.o.g. that each $\paramn \in \param$ is normalized and have used an eigenvector
normalization constrain with an arbitrary but fixed vector~$\mathbf{e}_{\star}$. 
Different modifications of the cost functions are explored in later Sections. 
Choosing $\mathbf{e}_{\star}=\mathbf{e}_k$ yields an $L^2$-normalized solution, i.e., 
\begin{equation}
    \int_{\Omega_\param} \mathbf{E}_k \cdot \mathbf{E}_k \,\mathrm{d}\mathbf{x}=1
\end{equation}
and may seem therefore natural but complicates the computation of the derivative \cite{Jorkowski_2018aa,Dailey_1989aa}.

For a gradient-based approach, we need an expression for the derivative of the objective function with respect to the optimization variable~$\param$, i.e., 
\begin{equation}
    \ddp g(\lambda(\param)) = - \left(\lambda_\mathrm{ref}- \lambda(\param) \right) \cdot \ddp \lambda(\param).
	\label{eq:diff_cost-fct}
\end{equation}
If this is not available, e.g., because the derivative of the eigenvalue is not computable in closed-form, the derivative of the objective function can directly be approximated using finite differences.
This variant is for example implemented in MATLAB\textsuperscript{\textregistered}'s nonlinear programming solver \texttt{fmincon} and is used unless the user provides their own derivative of the objective function when calling the optimizer. 

Alternatively, in order to determine the derivative of the eigenvalue, we first differentiate the generalized eigenvalue problem of formulation~\eqref{eq:constraintGEVP} with respect to $\param$ as well as the eigenvector normalization constraint~\eqref{eq:constraintNormalEigvec}.
Then, by solving the linear system of equations 
\begin{equation}
\begin{aligned}
	\begin{bmatrix}
		\stiff(\param) - \lambda_k(\param) \mass(\param)  &\! -\mass(\param)\mathbf{e}_k(\param) \\
		\mathbf{e}_{\star}^\top\mass(\param) &0
	\end{bmatrix}
	\begin{bmatrix}
		\ddp \mathbf{e}_k(\param)
		\\
		\ddp \lambda_k(\param)
	\end{bmatrix}=\\
	\begin{bmatrix}
	-\ddp \stiff(\param)\mathbf{e}_k(\param)+\lambda_k(\param)\ddp \mass(\param)\mathbf{e}_k(\param) \\
	-\mathbf{e}_{\star}^\top \ddp\mass(\param)\mathbf{e}_k(\param)
	\end{bmatrix},
 \end{aligned}
	\label{eq:eigprobderiv}
\end{equation}
we obtain the first-order derivative $\ddp \lambda_k(\param)$ of the eigenvalue and can calculate~\eqref{eq:diff_cost-fct}.
The derivatives of the system matrices can again be approximated via finite differences.
However, we want to employ closed-form derivatives whose formulation relies on the exact representation of the geometries using spline-based computer-aided design (CAD) basis functions.
Therefore, we choose Isogeometric Analysis for the discretization of the problem as explained in the following.

\subsection{Isogeometric Analysis} \label{iga}
In order to spatially discretize the computational domain, it can be approximately divided into simple (polynomially curved) shapes like tetrahedra, e.g., in the classical \textit{Finite Element Method} (FEM)~\cite{Monk_2003aa}.
However, if we use the same basis functions for the model as used in the construction of the CAD geometry, namely \textit{B-splines} and \textit{non-uniform rational B-splines} (NURBS), no geometry modeling-related error is introduced. 
These are chosen in the \textit{Isogeometric Analysis} (IGA), which we use for the discretization of our problem~\cite{Cottrell_2009aa, Vazquez_2010aa}.
The NURBS are obtained from the simpler \textit{B-splines}, which are implicitly defined by the knot vector
\begin{equation}
    \Xi=\left\{\xi_{1}, \xi_{2},\dots, \xi_{{n_\mathrm{dof}}+\degr+1}\right\}
\end{equation}
where $\degr$ is the desired polynomial order of the resulting B-splines, and ${n_\mathrm{dof}}$ is the number of basis functions used. 
B-spline basis functions are defined recursively, starting with the piece-wise constant functions ($\degr = 0$)
\begin{equation}
    N_{i,0}\left(\xi\right)=
    \begin{cases}
    1,&\mathrm{if}\hspace{0.2cm}\xi_{i}\leq\xi<\xi_{i+1},\\
    0,&\mathrm{otherwise.}
    \end{cases}
\end{equation}
For $\degr=1,2,\dots,$ they are defined by
\begin{equation}
    N_{i,\degr}\left(\xi\right)=\frac{\xi-\xi_{i}}{\xi_{i+\degr}-\xi_{i}}N_{i,\degr-1}\left(\xi\right)+\frac{\xi_{i+\degr+1}-\xi}{\xi_{i+\degr+1}-\xi_{i+1}}N_{i+1,\degr-1}\left(\xi\right) \mathrm{.}
\end{equation}
B-spline curves of degree $\degr$ can then be written as
\begin{equation}
	\mathbf{B}_\degr(\xi) = \sum_{i=1}^{n_\mathrm{dof}} N_{i, \degr}(\xi) \mathbf{P}_i \,\mathrm{, }
\end{equation}
where $\mathbf{P}_i$ is the $i$-th control point. 
By introducing weights $w_i$ to the basis functions, allowing non-uniform (i.e., not evenly spaced) knot vectors, and dividing by the weighted sum of all basis functions, we can represent more general curved boundaries. NURBS basis functions read
\begin{equation}
    R_{i, \degr}\left(\xi\right)=\frac{N_{i,\degr}\left(\xi\right)w_{i}}{\sum_{j=1}^{n_\mathrm{dof}}N_{j,\degr}\left(\xi\right) w_j} \,\mathrm{, }
\end{equation}
and the resulting curve can be written as
\begin{equation}
	\mathbf{C}_\degr (\xi) = \sum_{i=1}^{n_\mathrm{dof}} R_{i, \degr} (\xi) \mathbf{P}_i \,\mathrm{.}
\end{equation}
In addition to modification of the control points and the knot vector, we can now also influence the weights to change the overall appearance of a curve. Increasing the weight of a control point moves the curve closer to that point, decreasing it moves the curve further away. As we are now dealing with a rational function, the curvature constraints of polynomial basis functions no longer apply, enabling the description of more general geometries. 
More details on the discretization of the problem can be found, e.g., in~\cite{Ziegler_2023ab}.
The concept of gluing multiple patches to a \textit{multipatch} geometry allows for the representation of topological complex geometries, e.g. with holes.
To keep our presentation brief, we omit a detailed explanation and refer the reader, e.g., to~\cite{Cottrell_2009aa}.

\section{Sensitivities} \label{sec:iga_matrix_diff}
Based on the spline representation of the geometry used in IGA, we can compute the derivatives of the eigenpair with respect to the control points in closed-form. 
As demonstrated above, this requires the derivative of the system matrices.
For the Maxwell eigenvalue problem, the derivatives were introduced in~\cite{Ziegler_2023ab}.
Here, we will recall the method briefly and consider one direction~$\paramn \in \param$ in our derivations. 
For easier readability, we suppress the subscript $\nvar$ in this section.

\begin{figure}
\centering
\includegraphics[]{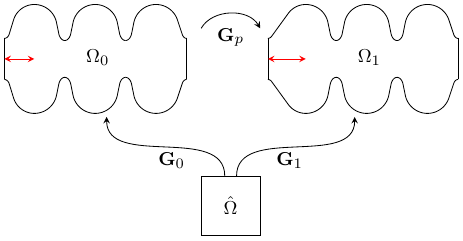}
\caption{Mapping~$\mathbf{G}_\paramnabbrev$ from an undeformed domain~$\Omega_0$ to the domain~$\Omega_1$ with a (for display purposes excessively) increased first half-cell length (marked in red).}
	\label{fig:mappingsLength}
\end{figure}

Within IGA, the physical domain is represented by a mapping $\mathbf{G}(\hat{\Omega})$ from the reference domain $\hat{\Omega}$. 
Thus, two different geometries $\Omega_0$ and $\Omega_1$ are represented via their respective transformations $\mathbf{G}_0, \mathbf{G}_1$, c.f. Fig.~\ref{fig:mappingsLength}.
If we consider a deformation from the domain $\Omega_0$ to $\Omega_1$ in dependence of a deformation parameter $\paramnabbrev \in [0,1]$, then the intermediate, deformed geometry $\Omega_\paramnabbrev$ is constructed by
\begin{equation}
	\Omega_\paramnabbrev = \mathbf{G}_\paramnabbrev (\Omega_0) = \mathbf{G}_\paramnabbrev (\mathbf{G}_0 (\hat{\Omega})).
 \label{eq:omega_t}
\end{equation}
Here, we have used the transformation mapping $\mathbf{G}_\paramnabbrev: \Omega_0 \to \Omega_\paramnabbrev$ in the form
\begin{equation}
	\mathbf{G}_\paramnabbrev (\mathbf{x}) = \mathbf{x} + \paramnabbrev \mathbf{V}_\paramnabbrev(\mathbf{x}) \, , 
	\label{eg:mapGp}
\end{equation}
with a smooth displacement vector field $\mathbf{V}_\paramnabbrev$ along which we move the initial point $\mathbf{x} \in \Omega_0$ by $\paramnabbrev$. 

We can use this to replace the integration over $\Omega_\paramnabbrev$ required in \eqref{eq:system-matrices} by integration over $\Omega_0$. Computing the derivative with respect to $\paramnabbrev$ yields
\begin{equation}
	\begin{aligned}
		\ddt \stiff_{i,j}(\paramnabbrev) &= \int_{\Omega_0} \ddt \left[ \mathbf{J}_\stiff(\paramnabbrev) \right] \curl \mathbf{v}_i \cdot \curl \mathbf{v}_j \, \mathrm{d}\mathbf{x}  \, ,\\
		\ddt \mass_{i,j}(\paramnabbrev) &= \int_{\Omega_0} \ddt \left[ \mathbf{J}_\mass(\paramnabbrev) \right] \mathbf{v}_i \cdot \mathbf{v}_j \, \mathrm{d}\mathbf{x} \, ,
	\end{aligned}
	\label{eq:system-matrices-ddt}
\end{equation}
where the terms
\begin{equation}
	\begin{aligned}
		\mathbf{J}_\stiff(\paramnabbrev) &= \frac{1}{\det (\partial_\mathbf{x} \mathbf{G}_\paramnabbrev)} \partial_\mathbf{x} {\mathbf{G}_\paramnabbrev}^\top \partial_\mathbf{x} \mathbf{G}_\paramnabbrev\, , \\
		\mathbf{J}_\mass(\paramnabbrev) &= \det (\partial_\mathbf{x} \mathbf{G}_\paramnabbrev) \partial_\mathbf{x} {\mathbf{G}_\paramnabbrev}^{-1} \partial_\mathbf{x} {\mathbf{G}_\paramnabbrev}^{-\top} ,
	\end{aligned}
\end{equation}
ensure the curl-conserving transformations \cite{Monk_2003aa}. By $\partial_\mathbf{x}$ we denote the Jacobians with respect to $\mathbf{x} \in \Omega_0$. 
The derivatives of the terms $\mathbf{J_K}$ and $\mathbf{J_M}$ are computed via MATLAB\textsuperscript{\textregistered}'s Symbolic Toolbox. 
The code used for this procedure is publicly available at~\cite{Ziegler_github}.
Using these, the derivatives of the system matrices \eqref{eq:system-matrices-ddt} can then be computed in closed-form. 
From those, $\ddt \lambda(\paramnabbrev)$ and $\ddt \mathbf{e}(\paramnabbrev)$ are obtained by solving the equation system~\eqref{eq:eigprobderiv}. 

\subsection{Non-linear Parameter Dependence}\label{sec:nonlinearV}
For the computation of the sensitivities, it remains to determine the appropriate displacement vector field~$\mathbf{V}_\paramnabbrev$ to calculate~\eqref{eg:mapGp}.
Using IGA, the idea of the implementation of the shape derivatives is based on the parameterization of the shape deformation in terms of the control points, which may depend on further design parameters.
Hence, we can formulate the smooth displacement vector field as
\begin{equation}
    \mathbf{V}_\paramnabbrev (\mathbf{x})  = \sum_{i=1}^{\nvar_\mathrm{dof}}  \frac{\mathrm{d}\mathbf{P}_i(\paramnabbrev)}{\mathrm{d}\paramnabbrev} R_{i, \degr}(\mathbf{x})\, .
\end{equation}

When the relation between the parameter~$\paramnabbrev$ and the control mesh is linear, we can express the displacement vector field in a straightforward way via
\begin{align}
    \mathbf{V}_\paramnabbrev(\mathbf{x}) &= \sum_{i=1}^{\nvar_\mathrm{dof}} \left(\mathbf{P}_{1,i} - \mathbf{P}_{0,i} \right) R_{i, \degr}(\mathbf{x})\\
    &= \mathbf{G}_1 (\mathbf{G}_0^{-1}(\mathbf{x})) - \mathbf{x}\, ,
\end{align}
where $\P_0$ and $\P_1$ correspond to the control meshes of domains $\Omega_0$ and $\Omega_1$, respectively. 

When the control mesh depends nonlinearly on the parameter, extracting the explicit formulation of the displacement vector may be challenging.
In this case, we propose a first-order finite difference approximation of the parameter-to-control-point mapping
\begin{equation}
    \frac{\mathrm{d}\mathbf{P}_i(\paramnabbrev)}{\mathrm{d}\paramnabbrev} \approx \frac{\mathbf{P}_i(\paramnabbrev+\delta)-\mathbf{P}_i(\paramnabbrev)}{\delta} \, ,
    \label{eq:dP_lin}
\end{equation}
which provides exact derivatives in the limit case $\delta \rightarrow 0$, and still good approximations for sufficiently small $\delta$.

Unfortunately, this variant exhibits the same computational effort per gradient computation as classical finite differences on the objective function, i.e., as performed by \texttt{fmincon}.
For the current parameter set~$\param$ of the iteration of the optimizer, we need to evaluate the system at an additional point $\paramn+\delta_\nvar$ for each $\paramn \in \param$.
However, due to the improved quality of the approximation, this results in a significant efficiency improvement for the optimizer, as we will see in the numerical results in Section~\ref{numerics}.

\section{Mode Matching for Eigenvalue Crossings} \label{crossing}
\begin{figure}[htb]
	\centering
    \includegraphics[]{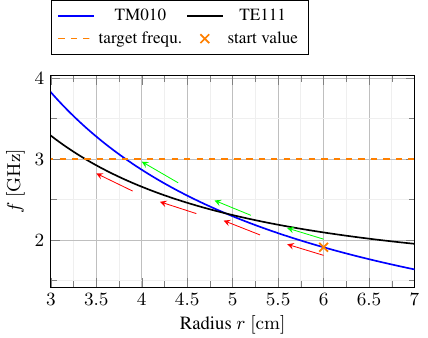}
	\caption{Plots of the analytical eigenfrequencies of the $\mathrm{TM}010$ and the $\mathrm{TE}111$ mode, which cross at $r = \SI{4.92}{\centi \meter}$.
    When optimizing the numerical frequency towards $f_\mathrm{ref} = \SI{3}{\giga \Hz}$ starting at $r_0 = \SI{6}{\centi \meter}$, the optimizer needs to move across this point. Applying the crossing detection and index correction method yields the frequencies of the correct mode at $r_\mathrm{opt} = \SI{3.82}{\centi\meter}$, as indicated by the green arrows. Without crossing detection, the optimizer switches to a different mode, as marked with the red arrows.}
	\label{fig:eigenvalue-crossing}
\end{figure}

The eigenvalue problem presented in \eqref{eq:eigenvalue-problem-discrete} has different modes $\mathbf{e}_k$ with distinct or partially identical eigenfrequencies $f_k$. 
We are, however, typically only interested in one single mode at a time, usually the accelerating one. 
Therefore, the problem of identifying the desired mode among all obtained solutions arises. 
If we know the appropriate index $k$ for the initial geometry, e.g., identified by the classification algorithm~\cite{Ziegler_2023aa}, the same index is only correct for deformations that are sufficiently small.
However, in general, crossings of the eigenfrequencies can occur and thus the index changes
since it is based on magnitude \eqref{eq:sort}. 
In this case, keeping the same index results in the wrong mode and thus the wrong frequency being selected for optimization. 
This phenomenon can be easily demonstrated for the pillbox cavity, for which the eigenfrequencies are known analytically~\cite{Jackson_1998aa} and in which the fundamental mode switches along variation of the radius.
Fig.~\ref{fig:eigenvalue-crossing} shows an example of this.
In the figure, we plotted the frequencies of the $\mathrm{TM}010$ and the $\mathrm{TE}111$ mode over a radius range.
For the considered cavity length of $\SI{10}{\centi \meter}$, the modes cross at $r = \SI{4.92}{\centi \meter}$.
If we always optimize the first eigenvalue, the considered mode will change when moving across this point.

In order to avoid erroneously switching to a different mode during the optimization, we employ an idea based on the method introduced in \cite{Jorkowski_2018aa}. 
This still requires us to know the correct index for the starting geometry. However, for each subsequent optimization step $i$, we determine the correlation coefficient
\begin{equation}
	\varphi_{i, k} = \frac{\mathbf{e}_{i, k}^\mathrm{H} \mathbf{M}_{i} \mathbf{e}_{i-1}}{\sqrt{\mathbf{e}_{i, k}^\mathrm{H} \mathbf{M}_i \mathbf{e}_{i, k} }\, \sqrt{\mathbf{e}_{i-1}^\mathrm{H} \mathbf{M}_i \mathbf{e}_{i-1}}}
\end{equation}
between the $k$-th eigenvector at iteration $i$, $\mathbf{e}_{i, k}$, and the known eigenvector $\mathbf{e}_{i-1}$ from the previous iteration $i-1$. The desired index $k$ then follows from the computed $\varphi_{i, k}$. Ideally, the correlation coefficients should attain the values
\begin{equation}
	\varphi_{i, k} = \begin{cases}
		1, & \text{if }k \text{ is the correct index, } \\
		0, & \text{otherwise.}
	\end{cases}
\end{equation}
Due to numerical inaccuracies, uncorrelated modes can still exhibit a correlation coefficient slightly above zero, and correlated modes can have a coefficient slightly below $1$. As those effects were usually negligible in our tests, we simply choose
\begin{equation}
	k = \arg \max_k \left( \varphi_{i, k} \right)
\end{equation} 
as the index to use in iteration $i$. 
We integrate this into the optimization procedure by evaluating the correlation coefficients in each call of the objective function.
If the index of the mode with the highest correlation to the mode of the previous iteration changes, we make the change and issue a warning.

\section{Applications} \label{numerics}
All implementations are carried out in MATLAB\textsuperscript{\textregistered} using the GeoPDEs package\cite{Vazquez_2016aa}.
We investigate two different application examples and for the evaluation of the performance we compare the gradients approximated using the \texttt{fmincon}-internal finite differences with our provided shape derivatives. 
All timings are measured on a standard laptop with Intel(R) Core(TM) i7-1065G7 \SI{1.30}{\giga \hertz} CPU and \SI{16}{\giga \byte} RAM and averaged over $10$ runs. Note, that the matrix assembly to compute the gradients is particularly slow because of the prototyping nature of the MATLAB\textsuperscript{\textregistered} implementation and the usage of quadrature-heavy splines.    

\subsection{Pillbox Cavity}
For the simple case of the cylindrical pillbox cavity, of course, no numerical optimization is actually needed and all relevant quantities are available from the analytical formulas. 
Nevertheless, we choose this benchmark example since here we can clearly illustrate the problem of the eigenvalue crossing as seen in Fig.~\ref{fig:eigenvalue-crossing}. 

We investigate the optimization with mode matching for the pillbox cavity with a length of $\SI{10}{\centi\meter}$, where we want to find the optimal radius for a given reference frequency using the optimization problem~\eqref{eq:opti_problem} with the objective function~\eqref{eq:cost-fct}. 
For our demonstration, we select a radius $r = \SI{6}{\centi\meter}$ as the initial start value and $f_\mathrm{ref} = \SI{3}{\giga \hertz}$ as the reference frequency to enforce the optimizer to move across the crossing of the frequencies of the $\mathrm{TM}010$ and the $\mathrm{TE}111$ mode at $r_\mathrm{cross} = \SI{4.92}{\centi\meter}$.
The initial value is marked with the orange cross in Fig.~\ref{fig:eigenvalue-crossing}.
We bound the admissible radius by $r_\mathrm{lower} = \SI{2}{\centi\meter}$ and $r_\mathrm{upper} = \SI{7}{\centi\meter}$ 
and discretize the geometry with second-degree splines which results in $540$ degrees of freedom.
Then, we start the optimization for the eigenmode with index one, which we have identified as the index of our eigenmode of interest, the $\mathrm{TM}010$ mode. 
Without a crossing detection, we would keep optimizing the first index, which results in a mode switch from the $\mathrm{TM}010$ to the $\mathrm{TE}111$ mode.
This path is indicated with the red arrows.
If we however check for crossings, we notice the switch and change the index.
In this case, we follow the path marked with the green arrows and optimize the desired mode.

\begin{table}[htb]
\renewcommand*{\arraystretch}{1.5}
\processtable{Solver statistics for the optimization of the numerical solution of the pillbox cavity. Measurement of CPU time averaged over $10$ runs.\label{tab:opti-pillbox}}
{\begin{tabular}{p{3cm}ccc}
\toprule
 & Finite Differences & \quad & Shape Derivatives \\ \midrule
$\frac{|f_\mathrm{opt}\!-\!f_\mathrm{ref}|}{f_\mathrm{ref}}$ & $7.05 \!\cdot\! 10^{-7}$ & & $4.16\!\cdot\! 10^{-6}$ \\
\#iterations & $7$ & & $5$ \\
function calls & $17$  & &  $10$  \\
CPU time & $\SI{28.42}{\s}$ & & $\SI{40.26}{\s}$  \\ \botrule
\end{tabular}}{}
\end{table}

Note, that in this example, the movement of the control points depends linearly on the radius variation.
Hence, the derivatives of the system matrices~\eqref{eq:system-matrices-ddt} are exact up to machine precision.
The computational results are given in Tab.~\ref{tab:opti-pillbox}. 
We observe, that the reference frequency was attained with a high level of consistency in both cases, i.e., with and without provided shape derivatives, as can be seen from the small relative errors of the optimal solutions compared to the reference frequency, which are both in the range of $10^{-6}$ to $10^{-7}$.
Using finite differences, the relative error was slightly smaller.
However, the optimization effort was lower when using the shape derivatives as fewer iterations and calls to the objective function were required, compared to using the \texttt{fmincon}-internal finite differences.
As expected, the CPU time is higher due to the effort for the assembly of the system matrices and their derivatives.

\subsection{TESLA Cavity} \label{tesla}
\begin{figure}[htb]
	\centering
    \includegraphics[]{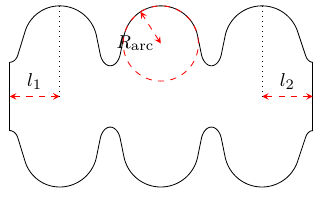}
	\caption{Geometry parameters of the TESLA cavity. We are concerned with tuning of the length of the first half-cell ($l_1$), the last half-cell ($l_2$), and the arc radius ($\Rtune$). For visualization reasons, here only one inner cell is depicted. In the 9-cell design, the inner cell type is repeated 6 more times.}
	\label{fig:tesla-params}
\end{figure}

As a more practical example, we apply the optimization to the TESLA cavity with $9$ cells. 
In this work, we are concerned with minor variations of the geometry parameters of the TESLA Test Facility (TTF) design shown in~\cite{Aune_2000aa}. Therefore, the deformations were small enough, that no eigenvalue crossing occurred.
We consider variations in three parameters following the numerical tuning procedure as described in \cite{Corno_2017ad}, in order to tune the cavity towards the desired resonant frequency and field flatness. 
The idea of the tuning process is based on the mechanical process in the manufacturing of the cavities but is simplified to a parameter variation since an exact mimicking of the mechanical process is complicated and not necessary~\cite{Corno_2017ad}. 
The tuned parameters are the length of the first half-cell, the length of the last half-cell, and the radius $\Rtune$ of the circular arc forming the equator.
They are shown in Fig. \ref{fig:tesla-params}.
We permit variations of up to $\pm 2 \, \mathrm{mm}$ for each parameter.
The reference frequency was chosen as $1.3\, \mathrm{GHz}$ for all following tests.

We will first formulate the objective function and then compare the results obtained after optimization with \texttt{fmincon}, once using the internal finite differences and once with provided shape derivatives.
Subsequently, we will investigate further extensions of the objective function.
We discretize the domain of the TESLA cavity with splines of degree two and thereby obtain $24,960$ degrees of freedom.
Should multiple solutions satisfy the accuracy requirement, we are interested in solutions requiring only small geometry changes. We thus use the objective function
\begin{equation}
	g(\param) = \left(f_\mathrm{ref} - f(\param) \right)^2 + s \left|\left| \param_\mathrm{diff} \right|\right|_2^2 \, \mathrm{, }
	\label{eq:obj-squared-err-t-penalty}
\end{equation}
which includes a penalty term for deviations $\param_\mathrm{diff}$ from the original geometry. 
The penalty factor $s$ has to be chosen such that the deviation from the reference frequency and the required geometry changes are balanced. 
If we require the error of the fundamental frequency $f_\mathrm{diff} = |f_\mathrm{ref} - f|$ to not exceed $10^5\si{\Hz}$ and want to keep $\param_\mathrm{diff}$ as small as possible, $s = 2\cdot10^{13}$ seems to be a reasonable choice in this case. 

\begin{table}[htb]
\renewcommand*{\arraystretch}{1.5}
\processtable{Nominal geometry parameters of the TESLA cavity: Length of the first half-cell ($l_1$), length of the last half-cell ($l_2$) and radius $\Rtune$ of the circular arc.
\label{tab:nominal-tesla}}
{\begin{tabular}{p{3cm}ccccc}
\toprule
& $l_1$ && $l_2$ && $\Rtune$ \\ 
\midrule
nominal & $\SI{56.0}{\milli \meter}$ && $\SI{57.0}{\milli \meter}$ &&  $\SI{42}{\milli \meter}$\\
\botrule
\end{tabular}}{}
\end{table}
We remark that for the variation of these three geometry parameters, the control points are moved in a non-linear way, as discussed in Section~\ref{sec:nonlinearV}.
Therefore, for the computation of the shape derivatives, we parametrize the control point displacement following the scheme of~\eqref{eq:dP_lin} with $\delta_\nvar=0.01\paramn$.
We then choose twelve sets of start values for which we run the optimization algorithm.
With the finite difference gradient computation approach, all tested start values lead to the same optimum. 
When using the shape derivatives to compute the gradient of the cost function, we obtain different values and select the best result.
The associated tuning parameters, different criteria for solution quality, as well as key performance figures, are shown in Tab.~\ref{tab:tuning-tesla} in the first two columns.
\begin{table}[htb]
\renewcommand*{\arraystretch}{1.5}
\processtable{Optimal values for the tuning parameters: Length of the first half-cell ($l_1$), length of the last half-cell ($l_2$) and radius $\Rtune$. Note that for the first two columns, $\eta_{1,\mathrm{opt}}$ and $\eta_{2,\mathrm{opt}}$ belong to the solutions which are optimal wrt. the frequency.
\label{tab:tuning-tesla}}
{\begin{tabular}{@{\extracolsep{\fill}}lrrr}\toprule
 & \multicolumn{2}{c}{Formulation \eqref{eq:obj-squared-err-t-penalty} with} & Formulation \eqref{eq:objfun2} for \\ 
 & Finite Differences & Shape Derivatives & Field Flatness\\ \midrule
$\Delta l_1 $ & $\SI{-0.02}{\milli \meter}$ & $\SI{0.07}{\milli \meter}$ &  $\SI{0.87}{\milli \meter}$ \\
$\Delta l_2 $ &  $\SI{-0.02}{\milli \meter}$ &  $\SI{-0.08}{\milli \meter}$  &  $\SI{1.01}{\milli \meter}$\\
$\Delta \Rtune $ & $\SI{0.71}{\milli \meter}$ &$\SI{0.69}{\milli \meter}$ &  $\SI{0.78}{\milli \meter}$\\
$f_\mathrm{opt}$ & $\SI{1.3001}{\giga \hertz}$ & $\SI{1.3002}{\giga \hertz}$ & $\SI{1.3001}{\giga \hertz}$ \\
$\frac{|f_\mathrm{opt}\!-\!f_\mathrm{ref}|}{f_\mathrm{ref}}$ & $6.25 \!\cdot\! 10^{-5}$ & $1.25 \!\cdot\! 10^{-4}$ & $7.11 \!\cdot\! 10^{-5}$ \\
$||\param_\text{diff}||_2$ & $0.18$ & $0.25$ &  $0.33$ \\
$\eta_{1,\mathrm{opt}}$ & $0.5559$ & $0.5247$ & $0.9808$ \\
$\eta_{2,\mathrm{opt}}$ & $0.8373$ & $0.8350$ & $0.9919$ \\
mean \#iterations & $10.75$ & $4.58$  &  $12.17$\\
mean function calls & $58.75$  &  $29.67$   &  $63.67$\\
mean CPU time & $\SI{2,080}{\s}$ &  $\SI{8,985}{\s}$  &  $\SI{2,277}{\s}$\\ \botrule
\end{tabular}}{}
\end{table}
One immediate observation is that using shape derivatives increases the used run time significantly. 
As we are dealing with a three-dimensional parameter space, explicitly computing the gradient requires the computation of three different derivatives in the respective parameter directions. 
When using the shape derivatives, in each iteration step, we thus have to assemble three deformed geometries in addition to the current geometry, before then computing three separate shape derivatives. 
On the other hand, the shape derivatives reduce iterations and function evaluations by $57\%$ and $50\%$, respectively.
Hence, the shape derivatives provide a significant efficiency increase for the optimizer. 
We note, that more efficient computation of the derivatives, e.g., computing the three required derivatives in parallel, could reduce the required computation time and is subject to further investigation.

\subsection{Field Flatness in TESLA Cavity} \label{flatness}
\begin{figure}[htb]
    \centering
    \includegraphics[]{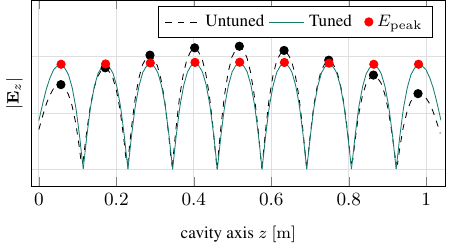}
    \caption{Magnitude of the longitudinal electric field component $|\mathbf{E}_z|$ on the central cavity axis. The peak values in each cell, ${E}_\mathrm{peak}$, are marked with dots. In the untuned cavity, the amplitudes vary strongly (marked in black). After tuning, the peak values are on a very similar level (marked in red).}
    \label{fig:fieldFlatness}
\end{figure}

The dynamics of the particle beam are affected by the electric field. Errors in phase and amplitude of the electric field cause beam degradation and losses \cite{Corno_2017ad,Edwards_1995aa} and the accelerating voltage should be maximized. Therefore, the tuning parameters need to be optimized in such a way that the amplitude of the accelerating electric field is the same in each cavity cell. 
Hence, we employ the field flatness criteria
\begin{equation}
	\eta_{1}(\textbf{p}) = 1 - \frac{ (\max_j |E_{\mathrm{peak},j}|-\min_j |E_{\mathrm{peak},j}|)}{\mathbb{E}(|E_{\mathrm{peak},j})|}
\end{equation}
and
\begin{equation}
\eta_{2}(\textbf{p}) = 1- \frac{\mathrm{std}(E_{\mathrm{peak},j})}{\mathbb{E}(|E_{\mathrm{peak},j}|)}
\end{equation}
introduced in \cite{Corno_2017ad}, where by $\mathrm{std}$ and $\mathbb{E}$, we denote the standard deviation and the expected value, respectively.
These criteria therefore provide a measure for an even distribution of the electric field peaks~$E_{\mathrm{peak}}$ along the axis of the cells. 
To keep the field quality and as such the beam quality within acceptable limits, $\eta_1, \eta_2 \geqq 0.95$ is typically required for a well tuned cavity \cite{Corno_2017ad}. 
The optimal solution found in the last section exhibits field flatness criteria of $\eta_1 = 0.5559$ and $\eta_2 = 0.8373$.
To improve these results, we combine the former objective function (\ref{eq:obj-squared-err-t-penalty}) with the quality characteristics for field flatness, i.e., 
\begin{equation}
\begin{aligned}
	g(\param) =&\; \bigr(1 - \eta_1(\textbf{p})\bigl) + \bigl(1-\eta_2(\textbf{p})\bigr) \\
            & + \alpha  \left(f_\mathrm{ref} - f(\param) \right)^2 + \beta \left|\left| \param_\mathrm{diff} \right|\right|^2 
 \end{aligned}
 \label{eq:objfun2}
\end{equation}
with $\alpha= 10^{-15}\;\si{\hertz}^{-2}$ and $\beta=\frac{2}{3} \cdot 10^{-2}$, which helps us to achieve an error within the accelerating frequency of $f_\mathrm{diff} = |f_\mathrm{ref} - f|  \leqq 10^{5}$. 
The improvement of the field flatness is illustrated in Fig.~\ref{fig:fieldFlatness}, where we plotted the amplitude of the longitudinal component of the electric field strength~$|\mathbf{E}_z|$.
We evaluated the magnitude in the cell centers to compute the field flatness criteria and marked these points with dots. 
In black, we indicate the field magnitude after optimization with respect to formulation~\eqref{eq:obj-squared-err-t-penalty}, which shows a poor field quality.
The results after optimization are marked in green and red.

The strongly improved field quality and much more even distribution of the field can also be seen from the much higher values of $\eta_{1,\mathrm{opt}}$ and $\eta_{2,\mathrm{opt}}$ in the right column of Tab.~\ref{tab:tuning-tesla}. 
After optimization with respect to the field quality, values of $0.9808$ and $0.9919$, respectively, are attained.
Even higher values would be achievable if we relaxed or dropped the constraint on the frequency and/or the penalty term, which keeps the deviations from the design small.
We can also make further observations.
Firstly, and also unsurprisingly, we note that for achieving the desired field flatness, we require a much larger extent of the geometry deformations, especially for the lengths of the half-cells.
The further computational statistics are slightly higher but in a comparable range to the ones obtained with the first formulation and finite differences, such as the relative deviation from the optimal frequency versus the reference frequency ($6.25 \!\cdot\! 10^{-5}$ vs. $7.11 \!\cdot\! 10^{-5}$), the mean number of required iterations ($10.75$ vs.  $12.17$), function calls ($58.75$  vs. $63.67$), and mean CPU time ($\SI{2,080}{\s}$ compared to  $\SI{2,277}{\s}$).

\section{Conclusion and Outlook}\label{conclusion}
In this work, we develop a gradient-descent based approach for the eigenvalue optimization using IGA.
The optimization is enhanced by using mode matching and shape derivatives which reduce the computational effort significantly, in some cases by over $50\%$.
In our experiments, the computational time, however, is elevated due to the prototype nature of the code. 
Albeit having demonstrated the approach for specific types of cavities, the algorithm can be used for arbitrary geometries.
This approach can easily be adapted to also be used in the design of new cavities, taking into account further design goals or allowing for free-form shape deformations.

\section{Acknowledgments}\label{sec11}

This work is supported by the Graduate School CE within the Centre for Computational Engineering at TU Darmstadt. 
We thank Peter Gangl for the fruitful discussions.

\bibliographystyle{iet}

\end{document}